\documentclass[graphicx, showkeys, aps, pre, groupedaddress, superscriptaddress]{revtex4-1}

\usepackage{graphicx}
\usepackage{dcolumn}
\usepackage{bm}
\usepackage[usenames]{color}
\usepackage{amsmath}
\usepackage{epstopdf}

\usepackage{ulem}

\begin{document}

\title{Thermal Marangoni-driven Dynamics of Spinning Liquid Films}

\author{Joshua A. Dijksman}
\affiliation{Physical Chemistry and Soft Matter, Wageningen University \& Research, Wageningen, The Netherlands}

\author{Shomeek Mukhopadhyay}
\affiliation{Department of Mechanical Engineering and Materials Science, Yale University, New Haven, CT 06511, USA}

\author{Robert P. Behringer}
\affiliation{Dept. of Physics \& Center for Nonlinear and Complex Systems, Duke University,  Box 90305, Durham, NC 27708-0305, USA}

\author{Thomas P. Witelski}
\affiliation{Department of Mathematics, Duke University, Box 90320,
Durham, NC 27708-0320, USA}

\date{\today}

\begin{abstract} 
Spin coating of thin films of viscous liquids on a rotating substrate is a core technological component of semiconductor microchip fabrication. The thinning dynamics is influenced by many physical processes. Specifically temperature gradients affect thin liquid films through their influence on the local fluid surface tension. We show here experimentally and numerically that adding a static temperature gradient has a significant effect on the equilibrium film thickness and height profile reached in spin coating. Our results suggest that thermal gradients can be used to control film height profile dynamics.
\end{abstract}

 \maketitle

\section{Introduction}
Thin layers of fluids on solid substrates display surprisingly rich dynamics, due to the interplay of a variety of forces at many lengthscales~\cite{simpson1982, 1985gennes, oron1997, craster2009, 2009bonn}. Much progress has been achieved on the study of thin fluid films. For low Reynolds number flows, they are well characterized by the lubrication approximation of the Navier-Stokes equation. This elegant approximate formalism allows for tractable analysis of a wide range of fluid dynamics problems on many lengthscales, such as liquids spreading on flat surfaces~\cite{ehrhard1991}, inclined surfaces~\cite{huppert1982}, 
convergent viscous gravity currents~\cite{dijksman2015},
spin coating applications~\cite{schwartz2004, wu2006}, flow of granular suspensions~\cite{ancey2013} and geophysical~\cite{balmforth2000} contexts, with ``thin'' here meaning that the height $h$ of the film is small with relative to the typical lateral lengthscale. The spatiotemporal evolution of the height field $h(x,y,t)$ is of a very general form, essentially a nonlinear conservation equation, describing how a profile in $h$  evolves in time. The evolution is driven by various forces, including gravitational, surface tension and, in a rotating system, centrifugal forces. If all the forces are sustained at constant levels, the film may approach an steady state profile.

\begin{figure}[tbp]
\includegraphics[width=14cm]{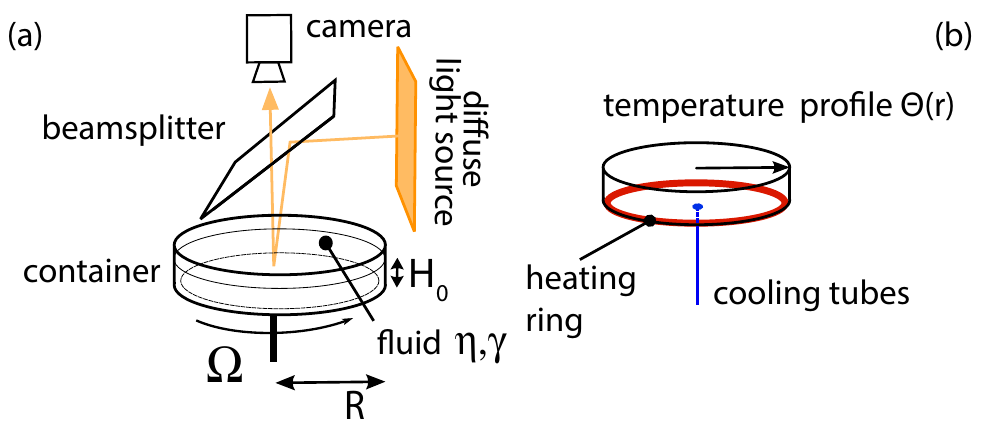}
\caption{\label{fig:setup_mech} (a) Schematic drawing of the container with the interferometry setup and all the relevant parameters: the rotation speed $\Omega$ in radians per second, the initial filling height $H_0$, the radius of the container $R$, the dynamic viscosity and surface tension of the fluid, $\eta, \gamma$ respectively. (b) the location of the heating element underneath the container, and double-walled rotating axis that doubles as a cooling tube.
}
\end{figure}

In a rotating container, the free-surface of a fluid will develop a parabolic free-surface profile to balance gravitational pressure and centrifugal forces with the amplitude of the profile increasing with the rotation rate~\cite{linden1984,lubarda2013,dijksman2015,bostwick2017}. For sufficiently large rotation rates, $\Omega> \Omega_c$, the profile will be truncated by the bottom of the container \cite{linden1984, dijksman2015}. Related flows are observed in other studies using a stationary container with a rotating bottom plate~\cite{bergmann2011,tophoj2013}. For a fluid that wets the container walls, a thin film of fluid will remain in the center of the container, we call this the central thin film (CTF), whose radius depends on the rotation rate~\cite{dijksman2015}.  Away from the center, the height profile remains parabolic. After sufficient time, the full height profile $h$ will converge to an equilibrium, due to conservation of mass in the container. 

In the absence of other influences, the CTF will be of nearly uniform thickness. The thinning dynamics of the film is described in the classic work of Emslie, Bonner and Peck~\citep{Emslie1958} (EBP), who considered the simplest case of centrifugal forcing of a viscous fluid. From the balance of viscous and centrifugal forcing, they derived that  $h(t) \propto t^{-1/2}$, a progressively slowing down thinning behavior. This behavior has been confirmed and expanded by many~\cite{acrivos1960, meyerhofer1978, flack1984, masahiro1987, kitamura2001} and is of fundamental importance in all spin coating techniques, for example in lithographic microchip production and in making the next generation of photovoltaics~\cite{delbos2012}.

Here we show that the addition of a surface tension gradient stress also known as Marangoni stress drives nontrivial spatio-temporal dynamics in a thin fluid film generated by spin coating. The inward Marangoni stress we create produces an accumulation of mass in the center of the film. We quantify how the thinning dynamics is affected by Marangoni stresses, and how the equilibrium profile is determined by the balance of Marangoni forcing with the various other forces acting on the thin film. We observe that Marangoni driving can even qualitatively change the structure of the surface at the edge of the CTF.

The paper first describes the experimental (Sec.~\ref{sec:exp}) and theoretical approach (Sec.~\ref{sec:theo}) used, and then presents the results in four main divisions: Sec.~\ref{sec:isothinning} shows how we can recover the classic EBP scaling dynamics; Sec.~\ref{sec:marathinning} describes how Marangoni forcing changes the EBP scaling. We also obtain results on the final equilibrium profile of the thin film spot after thinning has ceased: for the isothermal case, these results are described in Sec.~\ref{sec:isoequil}; the Marangoni effects on the equilibrium profile are described in Sec.~\ref{sec:maraequil}.

\begin{figure}
\includegraphics[width=16cm]{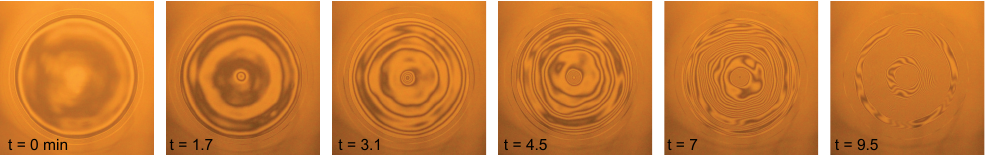}
\caption{\label{fig:maraqual} Fringe pattern of the CTF after steady rotation for 1~hour at $\Omega = 2\pi$ at isothermal conditions (leftmost panel). At $t=0$~min, heating/cooling is turned on. Even though the CTF has not equilibrated to its steady state, the Marangoni forcing immediately induces strong height variations in the CTF. Image crop measures about 4.6~cm in width.}
\end{figure}

\section{Experimental setup}\label{sec:exp}
The experimental system consists of an initially
uniform layer of fluid of thickness $\sim 3$~mm in a shallow cylindrical container -- see Fig.~\ref{fig:setup_mech}. The container is spun-up using a stepper motor to rotation speed $\Omega = 2\pi$~radians per second~(rps) unless otherwise noted. The container measures 13~cm in diameter and 2~cm in height. On the bottom of the container, a 4'' diameter silicon wafer (University Wafers) is placed; the wafer is fixed to the base through the deposition of a small ($\lesssim 1$~ml) amount of fluid between wafer and the bottom of the container. The suction force that keeps the wafer stuck to the container relies on the hydrodynamic drag on the thin film between wafer and container, and remains even after complete submersion of the wafer. The container is filled with a volume $V$ of fluid which gives an initial filling height $H_0 = V/(\pi R^2)$ with $R$ the radius of the container. We use polydimethylsiloxane (PDMS) for all experiments described in this work; this fluid completely wets the silicon wafer. The fluid wetting, combined with the fact that the thin films explored in this work are never thinner than several microns eliminates the necessity of extreme wafer cleanliness. The properties of PDMS are density $\rho = 965$~kg/m$^3$ and $\gamma = 0.02$~N/m. The thermal sensitivity of the surface tension for PDMS is $d\gamma/d\Theta \approx 6\times10^{-5} \mbox{N}/(^\circ\mbox{K$\cdot$m})$~\cite{1947fox, Bhatia1985}. We use a range of dynamic viscosities of 10-10000~mPa$\cdot$s; in all cases the viscosity and rotation rates used ensure that Coriolis forces do not play a role, since $\eta \gg \rho\Omega^2h$~\cite{Emslie1958}. The transparency of the PDMS and reflectivity of the silicon wafer allows for a laser-assisted alignment of the gravity-leveled fluid surface and the silicon wafer in the container, whose orientation can be tuned by set screws. Interferometry provides access to the spatio-temporal features of the thin film dynamics~\cite{toolan2012} -- see Fig~\ref{fig:setup_mech}a. Wafer illumination is provided with a uniform sodium light via a beam splitter. The spatial structure of the interference pattern of reflected and incoming light waves is recorded with a digital camera. In particular, height dynamics at any location can be measured through rate at which interferometric fringes evolve. The fringe succession rate can be measured with the varying intensity of the interference pattern, which we record by camera. The magnitude of the interference signal (the pixel values in the recording) is irrelevant, but the periodicity of the signal indicates the passing of fringes, which represent a height reduction (or increase) of known amplitude. Here we use this technique to measure the thin film dynamics in the center of the container at $r=0$. We will neglect the weak temperature dependence of the viscosity and the index of refraction (required for the fringe-based height measurements) of PDMS.
\par
In isothermal experiments, the container is uniformly heated to a temperature of 24$^{\circ}$C to fix the temperature dependent viscosity $\eta$ and surface tension $\gamma$ of the fluid. Temperature control is implemented by running water at a set temperature through the double-walled rotating axis. To establish Marangoni forcing, we cool the center of the container by running cooling water through the double-walled rotating axis while heating the outside ring of the container with a foil heater (Minco), positioned underneath the container (Fig.~\ref{fig:setup_mech}b). The Marangoni forcing has a very significant effect on $h(r)$. To show the qualitative effect of the initiation of heating on an almost flat CTF, see Fig.~\ref{fig:maraqual}. After isothermal spin-up of about 1~hour, we turn on Marangoni forcing. While the container is establishing its equilibrium radial temperature profile, we see strong evolution of the fringes in CTF region. 

The power for the foil heater is supplied through a slip ring (Moog) on the rotating axis. In the thermal gradient experiments, the level of cooling, set by the thermal bath that circulates the water (Neslab RTE7), is always run at maximum capacity. The maximum thermal gradient is then achieved at the largest heating power we can provide, which is 100W. The edge temperature at 100W heating is about 60$^{\circ}$C. For a complete description of the setup, see \cite{Mukhopadhyay2009}. This heating mechanism gives a azimuthally symmetric thermal profile in the bottom of the container, and hence on the silicon wafer, as shown in Fig.~\ref{fig:setup_IR}a. The temperature gradient is opposite in direction compared to the thermal profile considered in~\cite{dandapat1994}. 

Fig.~\ref{fig:setup_IR}b shows the temperature profile on a silicon wafer at maximum heating/cooling capacity in an empty container. We measure the profile with an infrared (IR) camera (FLIR A325). The infrared measurements require a known emmissivity for the substrate, and low reflectiveness of spurious infrared radiation into the camera. The measurements are thus performed with a layer of spray paint (Krylon flat white 1502) on the silicon wafer and in an empty container. We determined the infrared emissivity of the spray painted silicon wafer by calibrating the response at known temperature, similar to \cite{boreyko}. The spray painted wafer was subsequently placed in the container in the same way an uncoated wafer would be mounted in an experiment with an actual fluid present. The IR data is available only out to the edge of the silicon wafer, at approximately $0.8R$. 

\section{Governing model}\label{sec:theo}

\subsection{Temperature profile}
\par
\begin{figure}[!tbp]
\includegraphics[width=14cm]{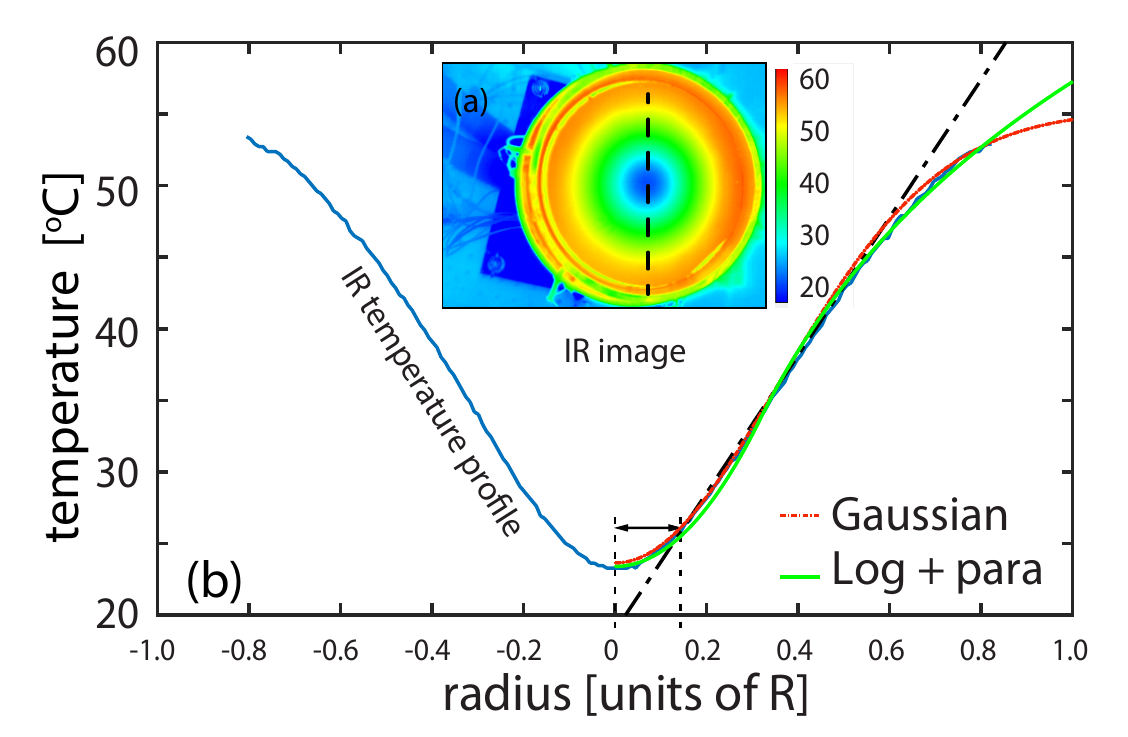}
\caption{\label{fig:setup_IR} (a) Infrared (IR) false color top view of the container showing the temperature profile of the base obtained at the largest thermal gradient possible. Color indicates temperature, ranging from $20^{\circ}$ to $60^{\circ}$K. The temperature profile on the dashed line is shown in (b) for the entire container, as a function of the radial position in units of the container radius $R$ (blue solid line). The black short-long dashes line indicates a linear fit with slope $7.4^{\circ}$K/cm. The actual thermal profile is well fitted to a smooth profile based on a Gaussian (red dash-dotted curve; see text) that approximates the piecewise-defined steady state solution (green solid curve). The arrow and dashed vertical line indicate the radius of the cooling tubes at the base of the container.}
\end{figure}

 
Due to the high thermal conductivity of the wafer, we expect the temperature profile to be maintained in the steady state set by the balance of the outer heating and central cooling. At the outer edge, the heating effectively sets the temperature at the boundary, $r=R$. On the interior, the temperature should satisfy the steady axisymmetric heat equation, with a heat sink, $q$, for the influence of the cooling,
To represent the idealized conditions, we write the steady state heat equation in cylindrical coordinates,
\begin{equation}
0 = \frac{\kappa}{r}\frac{\partial}{\partial r}\left(r\frac{\partial\Theta}{\partial r}\right) -q,
\end{equation}
in which $\kappa$ is the heat conductivity. We model the cooling as being a uniform constant value over a small inner region, $0\le r\le r_1$, and zero outside. 
Consequently, the temperature distribution can be described over the central and outer annular regions in terms of a parabolic profile and the classic axisymmetric steady state solution,
\begin{equation}
\Theta(r)= \Theta_0 + {q\over 4\kappa}
\begin{cases}
r^2 & 0\le r\le r_1\\
r_1^2 +2r_1^2\log(r/r_1) & r_1\le r\le R
\end{cases}
\label{v1temp}
\end{equation}
where $\Theta_0$ is the temperature at the origin. Fig.~\ref{fig:setup_IR} shows that
equation (\ref{v1temp}) matches the experimental profile well. Over a significant portion of the domain, the profile has a nearly uniform radial temperature gradient of about $7.4^\circ$K/cm. For PDMS\cite{1947fox, Bhatia1985}, this translates to a maximum surface tension gradient of about $4.4\times 10^{-2}$~N/m$^2$. 
It is important to note that it is not appropriate to approximate $\Theta(r)$ by a linear profile because this yields an unphysical gradient in the solution at the origin which can yield spurious behaviors.
\par
Fitting \eqref{v1temp} to the actual temperature profile data as shown in Fig.~\ref{fig:setup_IR}, we obtain an inner region where the parabolic profile applied with $r_1\approx 1.95$~cm ($r_1\approx 0.3R$, scaled relative to $R$). This radius is much larger than the width of the tubing $r_i\approx 0.17 R$ indicated in Fig~\ref{fig:setup_IR}b. We conclude that the assumption of uniform central cooling is not exactly satisfied and will replace \eqref{v1temp} with a qualitatively equivalent but less restrictive empirical profile.
\par 
The qualitative form of the temperature profile given by (\ref{v1temp}) should be mostly insensitive to variations in the properties of the central cooling, but we have not attempted to calibrate those values precisely. It will be convenient to replace this profile with an empirical fit to a single smooth function on $0\le r\le R$, given by
\begin{equation}
\Theta(r)=\Theta_0 +B\left(1 - \exp(-Cr^2)\right),
\label{v2temp}
\end{equation}
which has $\Theta'(0)=0$ at the origin. Here $B, C$ are dimensional fitting constants: $C$ relates to the effective width of the central cooling while $B$ scales with overall temperature rise to the outer edge of the container. The product $BC$ corresponds to the ratio of the source strength to conductivity from \eqref{v1temp}, $BC=q/(4\kappa)$, and the effective linear temperature gradient is given by the maximum slope, $\Theta'_{\max}=B\sqrt{2C/e}$.
This profile fits the experimental measurements well and is very close to (\ref{v1temp}) over most of the domain (see Fig.~\ref{fig:setup_IR}b). For our experimental setup, by fitting to the profile in Fig.~\ref{fig:setup_IR} we determined $C\approx 0.0935$~cm$^{-2}$.

\subsection{Lubrication model}
The time dependent film height $h(r,t)$ in the rotating container is described with a the time dependent axisymmetric lubrication approximation that includes surface tension, surface tension gradients, gravity, centrifugal force and disjoining pressure\cite{schwartz2004,wu2005,wu2006,wu2007,Mukhopadhyay2009}, 
\begin{equation}
- \frac{1}{3\eta r} \frac{\partial}{\partial r} \left\{ 
{\rho\Omega^2}r^2h^3 + 
\frac{3rh^2}{2}\frac{d\gamma}{dr}  - 
rh^3\frac{\partial}{\partial r} \left[ \rho g h - 
\frac{A}{h^3} \right]  
+ \gamma rh^3 \frac{\partial}{\partial r} \left[
\frac{1}{r} \frac{\partial}{\partial r}\left(r
\frac{\partial h}{\partial r} \right)\right] \right\} = 
\frac{\partial h}{\partial t} ~, 
\label{eq:filmfull}
\end{equation}
where $h$ is the height of the axisymmetric surface depending on the radial coordinate, $r$, and time, $t$. In the influence of the wetting properties of the container's base is given by the contribution of the disjoining pressure, $\Pi=A/h^3$, with negative Hamaker constant for complete wetting.
\par
To incorporate thermal Marangoni stresses, we use the temperature profile $\Theta(r)$ to write 
$$\frac{d\gamma}{dr} = \frac{d\gamma}{d\Theta} \frac{d\Theta}{dr}=\tau \frac{d\Theta}{dr}$$ 
with $\tau$ being a material parameter that captures the temperature dependence of the surface tension of PDMS. We assume a linear dependence of the surface tension $\gamma$ on temperature $\Theta$~\cite{ehrhard1991}; the literature suggests $\tau \approx 6\times 10^{-5}$~N/$^\circ$K$\cdot$m~\cite{1947fox, Bhatia1985}.
\par 
We nondimensionalize Eq.~\ref{eq:filmfull} with $h=H_0\tilde{h},  r=R\tilde{r}$ and set the timescale $T=\eta R^2/(\rho g H_0^3)$ based on the balance between viscous and gravity-driven effects. With these choices and after dropping the tildes on all nondimensionalized variables, the scaled equation is:
\begin{equation}
-\frac{1}{3r} \frac{\partial}{\partial r}\left\{
\mbox{Fr}^2\, r^2h^3 -
\mbox{Ma}\,\frac{3rh^2}{2}\phi(r)-
rh^3\frac{\partial h}{\partial r} -
\mbox{Ha}\,\frac{3r}{h}\frac{\partial h}{\partial r} 
+\frac{rh^3}{\mbox{Bo}} \frac{\partial}{\partial r}
 \left[
\frac{1}{r} \frac{\partial}{\partial r}\left(r
\frac{\partial h}{\partial r} \right)\right]
\right\}=\frac{\partial h}{\partial t}, 
\label{eq:filmfullscale}
\end{equation}
where the nondimensionalized temperature gradient function is
\begin{equation}
\phi(r)=2cr \exp(-c r^2)\qquad \mbox{with $c=3.95$},
\label{phiEqn}
\end{equation}
where $c=CR^2$ (analagous to a Damkohler or Thiele parameter for the dimensionless ratio of a reaction rate to a diffusivity) and other dimensionless parameters being
\begin{equation}
\mbox{Fr}^2={\Omega^2 R^2\over g H_0}\qquad 
\mbox{Ma}={\tau B\over \rho g H_0^2}\qquad
\mbox{Ha}={|A|\over \rho g H_0^4}\qquad
\mbox{Bo}={\rho g R^2\over \gamma},
\end{equation}
respectively: a rotational Froude number, a modified Marangoni number (the ratio of thermally driven surface tension gradients to gravity), a dimensionless Hamaker parameter, and a Bond number based on the size of the container.
\par
We use a second-order-accurate implicit finite difference scheme to solve the time dependent axisymmetric lubrication equation \eqref{eq:filmfullscale} subject to no-flux boundary conditions. In addition, we also solve for the steady-state profiles with a different quad-precision numerical code to give an independent check on the accuracy of the computational results for large times. We use values consistent with PDMS wherever they are constants; $\rho = 965$~kg/m$^3$ and $A = -7.6\times10^{-21}$J for a numerically convenient $\mbox{Ha} = 10^{-14}$.
\begin{figure}[!tbp]
\includegraphics[width=14cm]{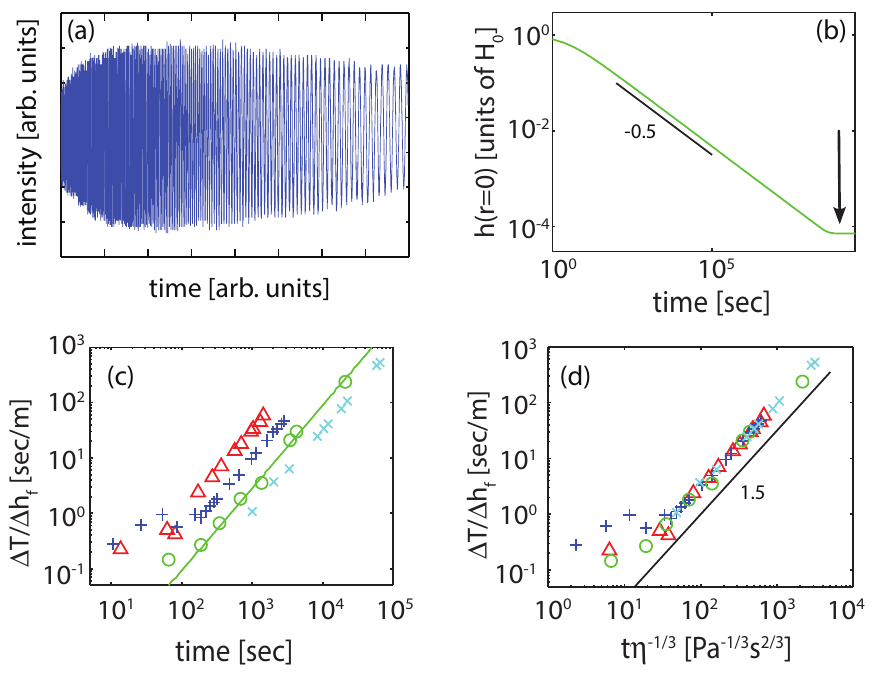}
\caption{\label{fig:thin_iso} (a) Typical interferometric signal at $r=0$ for a thinning experiment. The increasing period length of the intensity modulation in the interferometric signal signifies a decreasing thinning rate. (b) the result of numerically solving Eq.~\ref{eq:filmfull} with the parameters from a 1~Pa$\cdot$s spinning experiment. The solid black line has a slope corresponding to the EBP scaling of $t^{-1/2}$. The arrow points to the plateau in $h(r=0,t)$ when the equilibrium profile was reached. (c) Experimental data on the fringe succession time increase with time measured for various fluid viscosities, in units of mPa$\cdot$s: (10, red $\triangle$),  (100, indigo $+$),  (1000, green $\circ$),  (10000, light blue $\times$) at $\Omega = 2\pi$~rps. The green line corresponds to the data is the same as shown in panel (b), in the fringe succession time representation. (d) Rescaling the time axis with the EBP scaling of $\eta^{-1/3}$, we can collapse all data from panel (c). In the fringe succession representation, the expected EBP scaling corresponds to $t^{3/2}$ (see text) as observed (black line).}
\end{figure}

\section{Results}
\subsection{Isothermal thinning dynamics}\label{sec:isothinning}
We study the thinning dynamics for the system with $H_0 = 2.9$~mm, $\Omega = 2\pi$~rps, and PDMS oil with viscosities in the range $\eta = 10-10000$~mPa$\cdot$s with the fringe-passing technique described in Fig.~\ref{fig:thin_iso} and via solving Eq.~\ref{eq:filmfull}. Under steady rotation the thickness of the central film $h(r=0,t)$ will decrease until an equilibrium state is reached. Due to the thinning of the CTF during rotation, viscous forces increase progressively as the shear rates increase in the thinning layer, while the centrifugal force remains constant. The equilibrium solution is thus approached only very slowly. In our rotating container, we track the evolution of the thin film by recording the succession of fringes at $r=0$, the center of the container. Each successive fringe implies a thinning of the central thin film by $\Delta h_f = \lambda/2n = 210$~nm with $\lambda$ being the wavelength of the sodium light ($\lambda\approx 588$~nm) and $n$ the index of refraction of PDMS ($n\approx 1.4$). A typical intensity profile for the center of the container is shown in Fig.~\ref{fig:thin_iso}a: clearly the fringe succession period $\Delta T=t_{k+1}-t_k$, corresponding to
$h(0,t_{k+1})=h(0,t_k)-\Delta h_f$, grows over time. Solving Eq.~\ref{eq:filmfull} gives the expected EBP thinning dynamics with a scaling of $h(r=0,t) \propto t^{-1/2}$ in the CTF as shown in Fig.~\ref{fig:thin_iso}b. Eventually the numerical solution reaches a steady state profile in which the final equilibrium height is set by the disjoining pressure. Numerical data shown in aforementioned panel is obtained for the experimental conditions of a thinning experiment with a 1~Pa$\cdot$s fluid. We perform equivalent experiments for a range of different viscosities by counting the number of fringe successions as a function of time to obtain scaling of $h(r=0,t)$. All experiments are done at a rotation rate of $2\pi$~rps. 
\par
We cannot track all the passing fringes, as recording the entire experiment on video would result in prohibitively large data sets. Instead, we record short periods of the thinning process at several stages during the thinning process. Each recording is long enough to observe the length of a fringe passing period, so from every recording we can estimate the thinning rate. However, as we do not record the total number of passed fringes, we cannot get an accurate measure of the total change in the height profile. We can only record the thinning \emph{rate}. The minimum fringe passing time is of the order of several frames in the 30 frames per second video imaging, limiting the thinning rate measurements in the early stages of the thinning process. The manual recording of thinning dynamics make the thinning rate measurements irregularly spaced in time at later times. 
\par
Results are shown in Fig.~\ref{fig:thin_iso}c. The slowdown in thinning is clearly observed by the gradual increase in $\Delta T/\Delta h_f$. The numerical solution shown in Fig.~\ref{fig:thin_iso}b, from which we compute the derivative $\Delta T/\Delta h_f$ from the $h(r=0,t)$ data, also coincides with the corresponding experimental data without any free fitting parameters. Experimental data for fluids with different viscosities can be collapsed by rescaling the time axis with $\eta^{-1/3}$ as shown in Fig~\ref{fig:thin_iso}d, also consistent with EBP scaling. The rescaling shows clearly that the time between a fringe succession $\Delta T \propto t^{3/2}$, which implies that $h(t) \propto t^{-1/2}$, also consistent with the EBP scaling. These results show that our experimental and numerical methods in the rotating container geometry are effective in capturing the classic thinning dynamics. Interestingly, they indicate that the accumulation of fluid at the edge of the container during its rotation do not noticeably affect the EBP scaling for the thinning of the CTF. 
\begin{figure}[tbp]
\includegraphics[width=10cm]{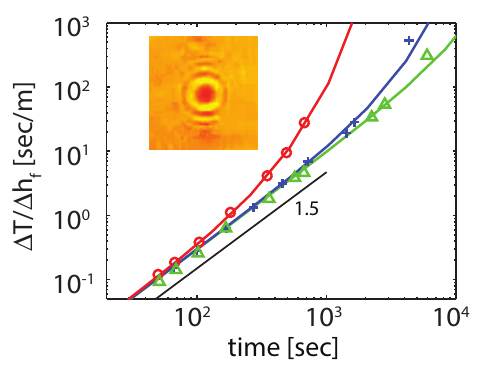}
\caption{\label{fig:thin_maraA}  Experimental (symbols) and numerical (solid curves) data on thin film height dynamics at $r=0$ during spinning with $\Omega = 2\pi$~rps and subject to different thermal gradients: 
($7.4^\circ$K/cm, red $\circ$), 
($6^\circ$K/cm, indigo $+$), 
($1.8^\circ$K/cm, green $\triangle$). 
For the numerics, we used $B=39,13,3.25$ corresponding to a maximum thermal gradient of: ($6^\circ$K/cm, red), ($2^\circ$K/cm, blue) and ($0.5^\circ$K/cm, green) to obtain best fits. Inset: digital image data, showing the circular fringes of the fluid hump in the center. Image width is approximately 3~mm. }
\end{figure}
\begin{figure}
\includegraphics[width=10cm]{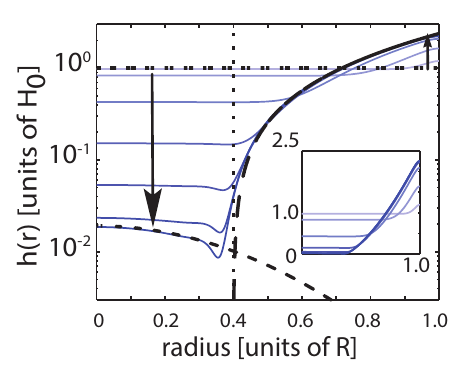}
\caption{\label{fig:thin_maraB} Numerical profiles  $h(r,t)$ corresponding to the $2^\circ$K/cm case from \ref{fig:thin_maraA} at times $0.01, 0.1, 1, 10,\cdots$, being effectively converged to a steady state by $t=10^5$s.
Arrows indicate the direction of motion of the evolving surface and dashed curves give predictions of the steady state profile to be constructed in sections   \ref{sec:isoequil} and 
\ref{sec:maraequil}.
The inset gives the profiles on a linear scale,
showing the semi-parabolic profile rapidly attained near the outer wall, to be described in
Section \ref{sec:isoequil}.  }
\end{figure}

\subsection{Marangoni effect in thinning dynamics}\label{sec:marathinning} 
We can now determine the effect of adding thermal Marangoni forces to spin coating applications. We probe the thinning dynamics for a $\eta = 100$~mPa$\cdot$s silicone oil spun at $\Omega = 2\pi$~rps with $H_0 = 2.9$~mm. In three different experiments, we provided an equilibrated, steady thermal gradient profiles of 7.4, 6 and 1.8$^\circ$K/cm. The thinning dynamics in representation $\Delta T/\Delta h_f$ are shown in Fig.~\ref{fig:thin_maraA}. The early time thinning behavior displays the classic EBP scaling with  $\Delta T/\Delta h_f \propto t^{3/2}$, corresponding to  $h \propto t^{-1/2}$. After some time however, the thinning dynamics slows down substantially, leading to what looks like a divergence of $\Delta T/\Delta h_f$. We check that our numerical simulations provide the same perspective. We can indeed quantitatively capture the experimental observations with the numerics -- see Fig.~\ref{fig:thin_maraA}. 
\par
Note that given our experimental settings, we allow for a small variation in the numerical value of the thermal stress gradient strength $b$, as we can only image the thermal profile at the base and have to assume a homogeneous temperature in the thin fluid layer. The divergence in the thinning time is accompanied by the appearance of a set of rings in the center of the rotating container (Fig.~\ref{fig:thin_maraA} inset). This suggests that the thermal Marangoni stress, which is directed towards the center of the container, draws fluid inward and serves to increase the height in the center of the container. The numerical simulations again confirm the experimental picture. In Fig.~\ref{fig:thin_maraB} we show computed $h(r)$ profiles at several times in the dynamics subject to a thermal gradient of $2^\circ$K/cm. After approximately $10^4$ seconds, the profile has effectively reached a steady shape with a central fluid hump. With the appearance of the hump, deviation with respect to EBP dynamics and indeed convergence to a finite-thickness steady state is expected. 
\par
\begin{figure}
\includegraphics[width=14cm]{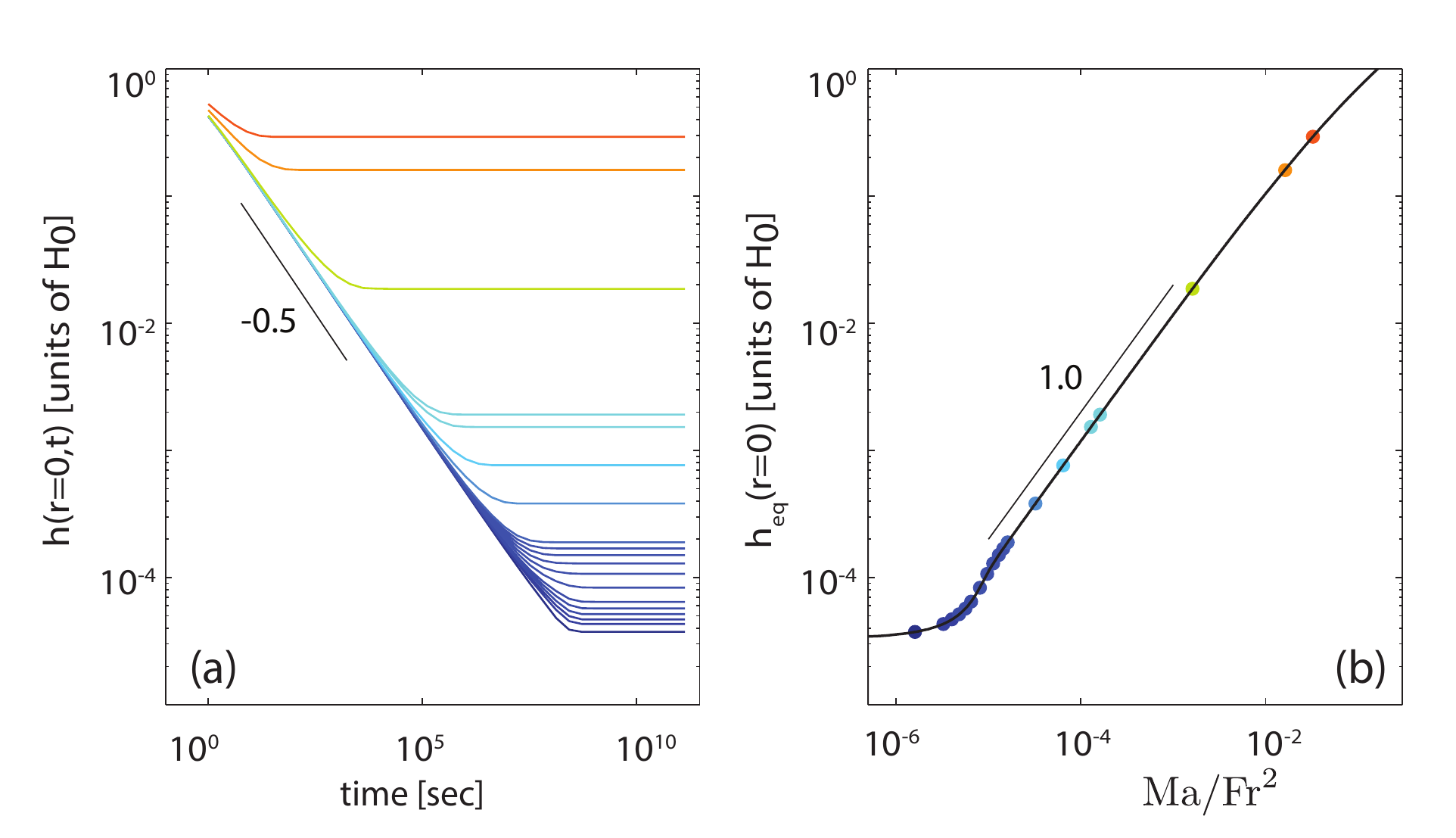}
\caption{\label{fig:longtime_mara} Numerical simulations for (a) $h(r = 0,t)$ for a range of Marangoni strengths; the color coding is the same as in (b). The standard EBP scaling of -0.5 is satisfied initially (black line, Eq.~\ref{hbarsol}); eventually an equilibrium height $h_{\mathrm{eq}}$ is reached as indicated by the plateaus, \eqref{hbarstar}. (b) the equilibrium heights $h_{\mathrm{eq}}(r = 0)$ for a range of Marangoni strengths $\mbox{Ma}/\mbox{Fr}^2$; the indicated linear scaling in $\mbox{Ma}$ is consistent with the prediction from
\eqref{hbarstar}.}
\end{figure}

The Marangoni effect on the thinning behavior can be understood from a simplified version of \eqref{eq:filmfullscale}. From Fig.~\ref{fig:thin_maraB} we observe that during most of the dynamics the height profile in the center of the container remains nearly flat,
$h(r,t)\approx \bar{h}(t)$. Therefore, capillarity can be neglected since the curvature will be small. We will also neglect disjoining pressure for short to moderate times while the CTF remaining relatively thick. Consequently, the
$r\to 0$ limit yields the leading order equation
\begin{equation}
{d\bar{h}\over dt} = -{2\over 3} \mbox{Fr}^2 \bar{h}^3 \left(1 - {3c\mbox{Ma} \over \mbox{Fr}^2 \, \bar{h}}\right),
\label{hbarODE}
\end{equation}
where $c$ is the parameter from \eqref{phiEqn} relating to the variance of the temperature profile about the origin. When $\bar{h}$ is relatively large, the factor in parentheses will be close to unity and the solution will be
\begin{equation}
\bar{h}(t)\sim \left(1+ {\textstyle{4\over 3}} \mbox{Fr}^2 \,t\,\right)^{-1/2},
\label{hbarsol}
\end{equation}
which corresponds to the EBP scaling, see Fig.~\ref{fig:longtime_mara}a. 
For longer times, as $\bar{h}$ becomes smaller,
the influence of the Marangoni stress is to slow the EBP thinning rate and to establish an equilibrium film thickness where Marangoni stresses and centrifugal forcing balance, 
\begin{equation}
\bar{h}_*= {3c\mbox{Ma} \over\mbox{Fr}^2},
\label{hbarstar}
\end{equation}
see Fig.~\ref{fig:longtime_mara}. Consequently,
Eqn \eqref{hbarsol} gives an estimate of the (dimensionless) time when a near-equilibrium thickness has been reached, $t_* = 3(h_*^{-2}-1)/(4\mbox{Fr}^2)$, for small Ma, yielding $\bar{h}_* \ll 1$, this
yields $t_*\sim (\mbox{Fr/Ma})^2/(12c^2)$. For our experimental settings at relatively large $\mbox{Ma}$, this yields dimensionless $t_*\sim 10^4$ and larger; as the time scale $T=\eta R^2/(\rho g H_0^3)$ is approximately 1.65, this equilibration estimate consistent with Fig.~\ref{fig:thin_maraA}. Since the timescales needed to explore the thinning dynamics become prohibitively large for smaller $\mbox{Ma}$, and experimental control of the temperature gradient is not ideal with smaller temperature gradients, we will explore the long time behavior only numerically.
\par
Note that Fig.~\ref{fig:longtime_mara}b shows deviations from the linear scaling with respect to $(\mbox{Ma/Fr}^2)$ for small $h$ (and very small Ma); we will see that this occurs when the disjoining pressure is no longer negligible compared to Ma and establishes a minimum thickness for the CTF layer.

\subsection{Isothermal equilibrium profiles}\label{sec:isoequil}
\par
To better frame the influences of the Marangoni stresses on the steady CTF profile, we first review behaviors for the isothermal free surface fluids in rotating containers \cite{linden1984}. Assuming $h=O(1)$, allowing us to neglect disjoining pressure effects, for 
small rotation rates the steady free surface will have a central depression that is paraboloidal, 
\begin{equation}
h(r) =1 +{\mbox{Fr}^2\over 2}\left(r^2 -{1\over 2}\right)
+ \mbox{Fr}^2\left({2\over \mbox{Bo}} -
{ I_0(r\sqrt{\mbox{Bo}}\,)\over \sqrt{\mbox{Bo}}\, I_1(\sqrt{\mbox{Bo}}\,)}\right).
\label{heq8}
\end{equation}
This result incorporates the contribution of surface tension through a term involving the ratio of modified Bessel functions $I_i$. Surface tension has a weak influence on the form of the solution, yielding a boundary layer of width $O(\sqrt{\mbox{Bo}})\to 0$ at the outer wall of the container to satisfy a contact angle condition, here taken to be $h'(1)=0$.
\par
For higher rotation rates \cite{linden1984}, a central bare spot will form,
\begin{equation}
h(r)\approx
\begin{cases}
0 & 0 \le r < r_c\\
{\textstyle {1\over 2}}\mbox{Fr}^2(r^2 -r_c^2) &
r_c < r< 1
\end{cases}
\label{eq9}
\end{equation}
where the radius of the bare spot is given by
\begin{equation}
r_c \approx \sqrt{1 -{2 \over \mbox{Fr}}}\ge 0\qquad \mbox{for $\mbox{Fr}\ge 2$.}
\label{eq13rc}
\end{equation}
Note that for large rotation rates, this shows that the fluid volume becomes forced into a narrow layer at the outer walls of the container, of width $1-r_c\sim 1/\mbox{Fr}=O(\Omega^{-2})\to 0$ as $\Omega\to\infty$. Similar expressions for a ``fluid hole'' were derived in \cite{bostwick2017}, but there the influence of gravity was neglected. This expression for the radius of the hole gives a very good estimate of the critical Froude number corresponding to the maximum rotation rate for the onset of formation of a hole, $\mbox{Fr}_c=2$, or equivalently, $\Omega_c=2\sqrt{gH_0}/R$.
\par
\begin{figure}
\includegraphics[width=3.5in]{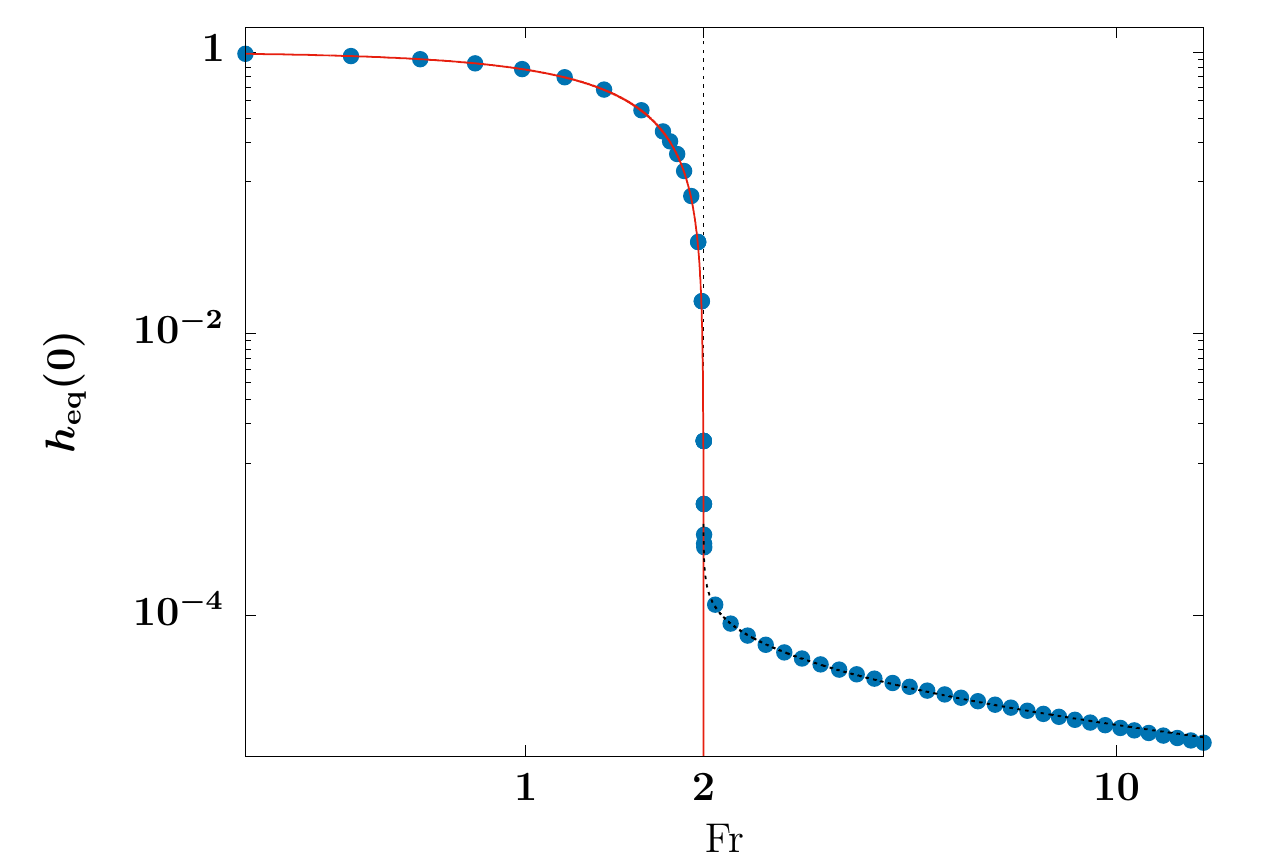}
\includegraphics[width=3.5in]{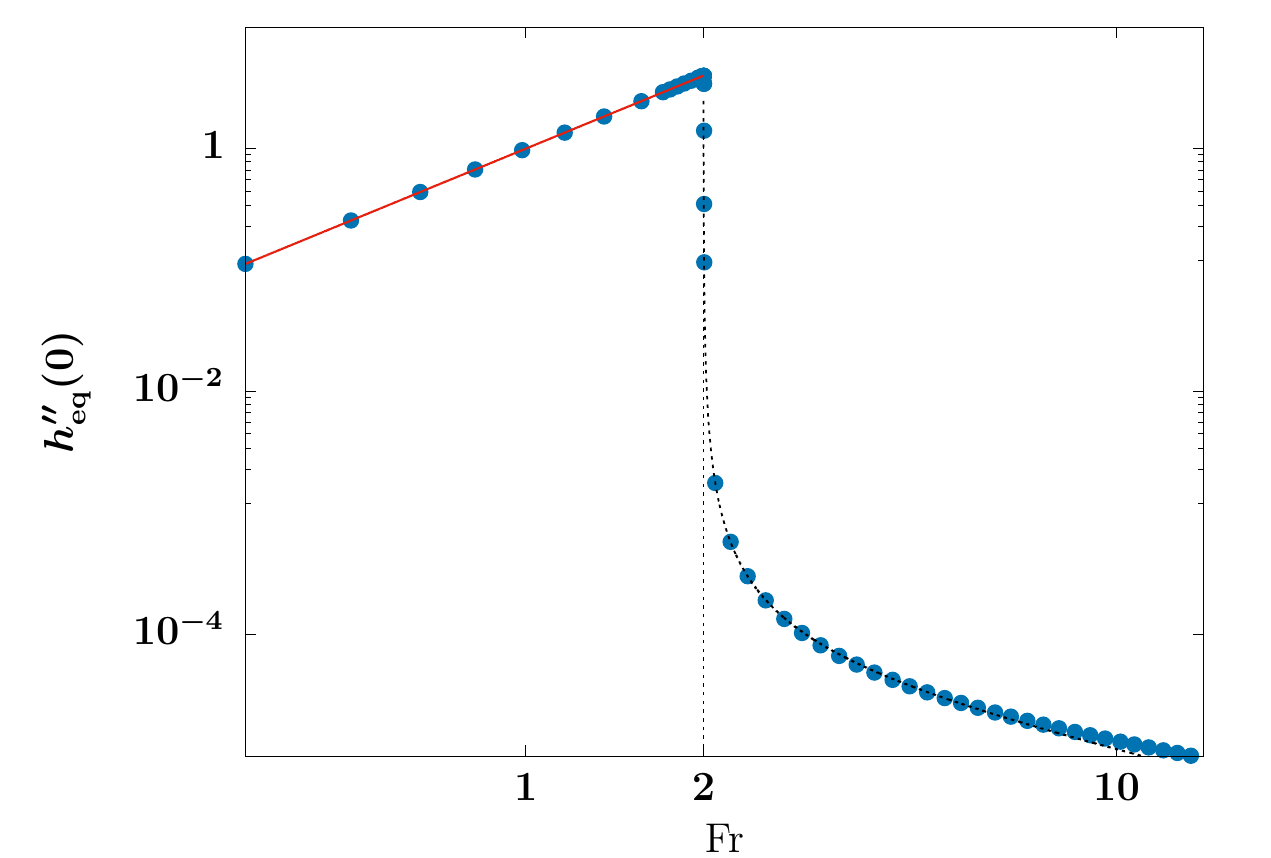}
\caption{Numerically computed properties of the isothermal equilibrium solutions of Eqn \eqref{eq:filmfullscale} on log-log plots: (a) Thickness of the film at the center, $h(r=0)$, showing excellent agreement with the predicted dependence on rotation rate, $h(0)=1-\mbox{Fr}^2/4$ for $\mbox{Fr}<2$ from \eqref{heq8} (red curve) and a fit to an empircal fit (dashed) for higher rotation rates, (b) The curvature at the center of the container, $h''(r=0)$, with analytically predicted behaviors for low rotation rate ($h''(0)=\mbox{Fr}^2$) and \eqref{eq15hpp} for larger
Fr.}
\label{fig:wm0hpp}
\end{figure}

In reality, on a completely wetting substrate, the ``bare spot'' will not be dry and will retain an adsorbed thin film due the intermolecular forces with the substrate, this describes our CTF. Neglecting the influences of capillarity and gravity for very thin films, by balancing the effects of the centrifugal effects with the disjoining pressure, we can obtain an approximate steady height profile for the CTF region,
\begin{equation}
\mbox{Fr}^2 r^2 h^3- \mbox{Ha} {3r\over h} {dh\over dr}=0\qquad \to\qquad 
h(r)= \left( h_0^{-3} - {\mbox{Fr}^2\over 2 \mbox{Ha}}r^2\right)^{-1/3},
\end{equation}
where $h_0=h(0)$ is the height at the origin. This solution can be used to
produce a scaling relation for the curvature of equilibrium solutions at $r=0$,
\begin{equation}
h''(0)= {\mbox{Fr}^2\over 3\mbox{Ha}}h(0)^4> 0.
\label{eq15hpp}
\end{equation}
The scaling for $h(0)$ for $\mbox{Fr}>2$ is not clear, based on the numerical simulations we have fit the data to an empirical relation like  $h(0)\propto (\mbox{Fr}-2)^{-0.25}/\ln(\mbox{Fr})^{0.72}$, see Fig.~\ref{fig:wm0hpp}a.
In summary, when Marangoni forcing is absent, the curvature of the film at the origin is positive for all rotation rates, but above the critical Froude number, the central curvature becomes much smaller and depends sensitively on the wetting properties of the substrate. While Fig.~\ref{fig:wm0hpp}a shows that the simplified prediction for the critical Froude number agrees very well with the full simulations, Fig.~\ref{fig:thin_maraB} shows that 
neglecting capillary effects and the disjoining pressure does affect the profile near the predicted CTF radius given by \eqref{eq13rc}.

\begin{figure}[tbp]
\includegraphics[width=6in]{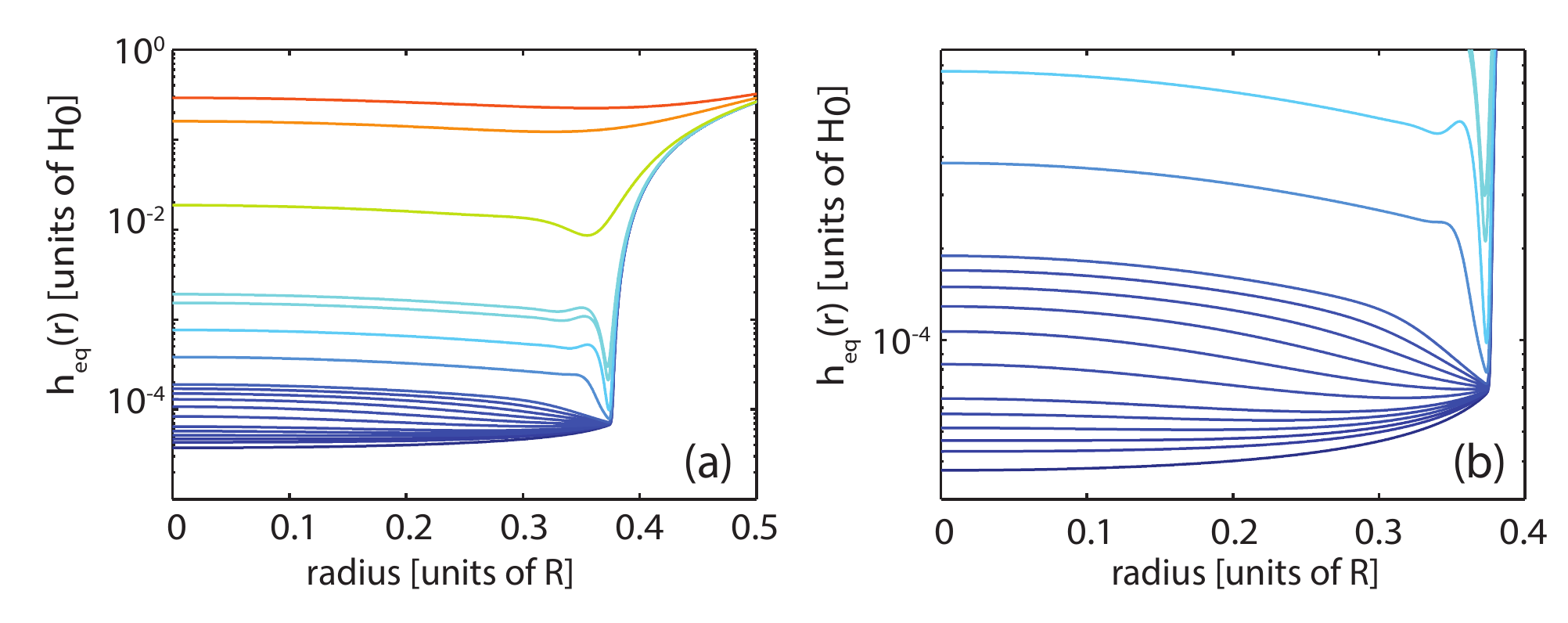}
\caption{\label{fig:profile_mara} (a) Computed $h_{\mathrm{eq}}(r)$ for a range of Marangoni strengths equivalent to Fig.~\ref{fig:longtime_mara}. The color scale is the same in all panels. (b) shows a close-up on the CTF region for small heights (and small Ma) from the profiles in panel (a). We observe that the curvature at $r=0$ changes sign as the limit $\mbox{Ma}\rightarrow 0$ is approached.}
\end{figure}

\subsection{Marangoni effects on the equilibrium profile}\label{sec:maraequil}
\par
To explore the equilibrium profile dynamics at finite Marangoni number, we again use the experimental values for all parameters in the numerical exploration and pick 100~mPas for viscosity and $2\pi$~rps for the spinning rate. Depending on the strength of the Marangoni stresses, we observe that the thinning in the center of the container initially follows the standard EBP scaling and reaches a minimum equilibrium thickness for the entire range of Marangoni strengths explored: see Fig.~\ref{fig:longtime_mara}a. 
\par
As described earlier, centrifugal effects scaled by the Froude number work to force fluid out of the central region, with the disjoining pressure and gravity opposing  this outflow. Thermocapillary effects due to the imposed temperature gradients will promote opposing inwards flows. To explore the full range of behaviors that can occur from difference balances of these effects, we use numerical simulations to compute the steady state solutions of \eqref{eq:filmfull} over a range of Marangoni numbers.
\par
Fig.~\ref{fig:profile_mara} shows steady profiles in the central thin film region for a range of Marangoni numbers, with other parameters fixed in the regime with $\mbox{Fr}>2$.  Fig.~\ref{fig:profile_mara}b shows that for very small Ma, the central thin film will have positive curvature; this is to be expected from the result \eqref{eq15hpp} for $\mbox{Ma}=0$. However,  we observe that for stronger thermal forcings, Marangoni stresses are sufficiently strong to draw in fluid to form a central ``hump'' with a local maximum, $h''(0)<0$.  
\par
When the film is thin smooth and slowly varying, for large Bond numbers we can neglect surface tension in the central region to approximate \eqref{eq:filmfull} by a first order equation for the steady CTF profile with no flux through the origin,
\begin{equation}
{dh\over dr} = \left(r\mbox{Fr}^2-{3\mbox{Ma}\over 2h} \phi(r)\right)\bigg/\left( 1+ {3\mbox{Ha}\over h^4}\right).
\label{eq16}
\end{equation}
This equation on $0\le r < r_c$ must be asymptotically matched to an interior layer at $r_c$ that captures capillary effects at the contact line and allows for
matching to the outer solution \eqref{eq9} on $r_c<r\le 1$. In general, determining the value of $h(0)$ will depend on this matching process, but we will show that over a range of larger Ma, a simpler solution can  be obtained.
\par
Fig.~\ref{fig:longtime_mara}b shows that for $\mbox{Ma}\to 0$, a minimum film thickness will be set as a function of Ha via the influence of the disjoining pressure. While the curvature of the CTF changes sign with Ma, the central height $h(0)$, is always monotone increasing with Ma. Eqn \eqref{eq16} gives a good approximation of the CTF profile up to a transitional range in Ma where surface tension starts to play a more important role in setting the structure of the film at $r_c$, see Fig.~\ref{fig:profile_mara}b.
\par
\begin{figure}
\includegraphics[width=6in]{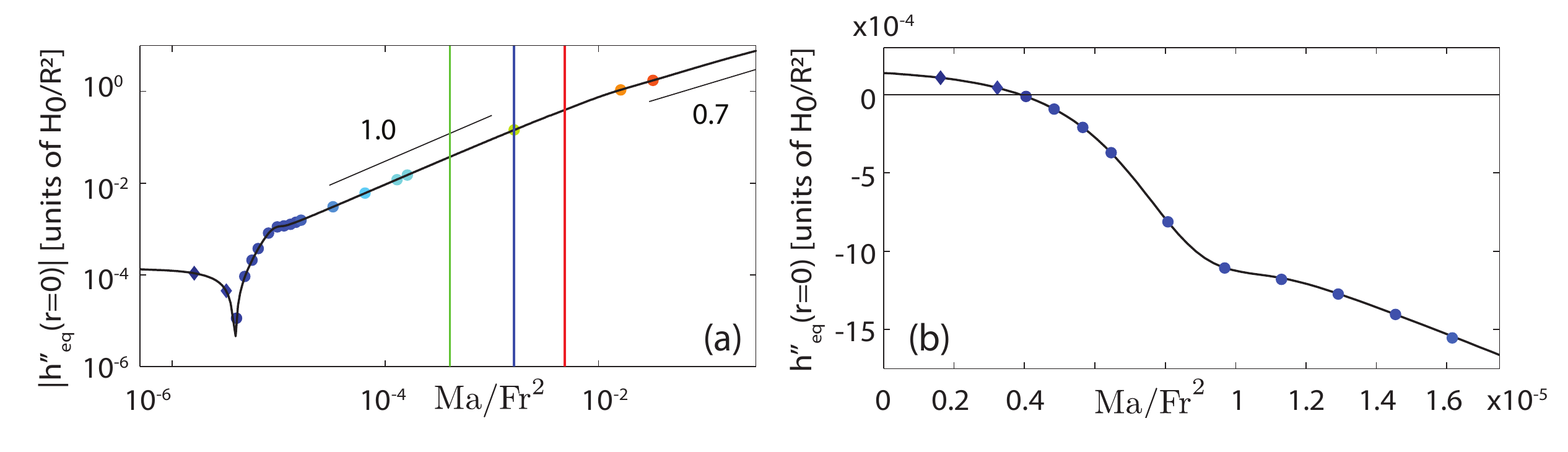}\caption{(a) The absolute value of the curvature of the steady state film height at the origin, $h''_{\mathrm{eq}}(r=0)$. Diamonds indicate positive value (a local minimum); filled circles represent negative curvature (a central hump) obtained from long-time runs of the dynamic problem \eqref{eq:filmfullscale}. The solid line was obtained by solving the steady state version of the equation over a range of Ma values. Scaling regimes are indicated with the thin solid lines and accompanying scaling exponents. The green, blue and red vertical line indicate the experiments at low, intermediate and high Ma respectively from Fig.~\ref{fig:thin_maraA}. (b) Zoomed-in view on a linear scale of the small Ma regime, showing how the curvature switches from positive to negative. }
\label{fig:Mahhpp}
\end{figure}

For Ma above this range, surface tension is still important locally at $r_c$, but in the CTF region, the centrifugal and thermocapillary influences dominate in \eqref{eq16} to balance and
give an explicit leading order estimate of the height profile in terms of the scaled gradient of the temperature profile,
\begin{equation}
h(r)={3\mbox{Ma}\over 2 \mbox{Fr}^2} {\phi(r)\over r} \qquad \mbox{on $0\le r< r_c$.}
\label{eq17}
\end{equation}
This well defined hump profile gives $h(0)\sim 3c(\mbox{Ma/Fr}^2)$ and
$h''(0)\sim 3c^2(\mbox{Ma/Fr}^2)$ yielding the linear scaling regimes seen in 
Figures~\ref{fig:longtime_mara}b and \ref{fig:Mahhpp}a. For even larger Marangoni numbers, this scaling ends when the thermal stresses are able to pull in the fluid from the outer region, to significantly degrade the semi-parabolic profile in \eqref{eq9}. 
\par
As suggested by the variation in forms of the height profiles for small Ma shown in Fig.~\ref{fig:profile_mara}b, Figure~\ref{fig:Mahhpp}b indicates that the central curvature has a nontrivial dependence on the system parameters in \eqref{eq16} and capillarity to yield the non-uniform behavior shown. In computations of the steady solutions with higher rotation rates, it was found that $h''(0)$ could become  
non-monotone with respect to $\mbox{Ma}$ at the transition of $\mbox{Ha}$ to $\mbox{Ma}$ dominated behavior occurring near Ma/Fr$^2\approx 10^{-5}$.

\section{Conclusions} We performed experiments and examined numerical solutions on thin film dynamics and steady state profiles in a rotating container in the presence of a thermal surface tension gradient force. We find that such thermal Marangoni forces can significantly affect the profile thickness and spatial height variations of the central thin film that develops at large enough rotation rates. Most notably, we find that the equilibration CTF height scales linearly with $\mbox{Ma}$. Once equilibrated, the CTF height profile follows the thermal profile. In the limit of small $\mbox{Ma}$, reaching equilibrium takes progressively longer and the steady state reached is set by competition between Marangoni and disjoining pressure. 

We foresee that this feature can be used in various applications. The dramatic changes for small Ma shown in Figures~\ref{fig:profile_mara} and \ref{fig:Mahhpp} will be observable in experimental fringe patterns for the CTF. We expect that this can developed further to yield a method for determining properties of the disjoining pressure (and characterize wetting properties of the substrate) by tuning the Marangoni number of a small range. The slow thinning dynamics need not be prohibitive: spin coating is done with rotation speeds that are orders of magnitude larger than used in this study, and the viscosity of the PDMS liquids (used for their low volatility) can also be orders of magnitude lower. 

The sensitivity of the spatiotemporal thin film dynamics to surface tension gradients can be of great interest in fields where functional nanometer thin films are produced with spin coating techniques~\cite{jiang2004}. Even thin film fluid deposition methods used in 3D printing can be improved with thermal gradient technology to design features smaller than a thickness of the fluid layer. If a temperature field is not the most natural control method, surface tension gradients can also be induced with other modes of
forcing like electric fields~\cite{electro2005} or  light~\cite{shin1999}. 

Fundamentally, there is also interest in exploring whether the competing centrifugal versus thermocapillarity influences can give rise to undercompressive shocks and fingering instabilities, as in the studies by Bertozzi and collaborators for planar thin films \cite{bertozzi1,bertozzi2,bertozzi3}.

\acknowledgements We thank Joshua Bostwick, Omar Matar, Howard Stone, Dominic Vella, Detlef Lohse and Jacco Snoeijer for stimulating discussions. Frans Leermakers helped with the numerical simulations and Chuan-Hua Chen and Jonathan Boreyko allowed and jhelped us use their infrared camera. This project was funded by NSF DMS0968252.

\bibliography{thintest}

\end{document}